\documentclass[12pt]{iopart}

\usepackage{graphicx,color,xcolor}
\usepackage[switch]{lineno}

\usepackage{mathrsfs}
\usepackage{cite}
\usepackage{iopams}

\newcommand{\GP}[1]{Gross-Pitaevskii }
\newcommand{\BdG}[1]{Bogoliubov-de Gennes}

\newcommand{\beq}{\begin{eqnarray}}
\newcommand{\eeq}{\end{eqnarray}}

\begin{document}

\title[Multiple period states of the superfluid Fermi gas in an optical lattice]{Multiple period states of the superfluid Fermi gas in an optical lattice}

\author{Sukjin Yoon$^{1,2}$, Franco Dalfovo$^{3}$, Takashi Nakatsukasa$^{4,5}$ and Gentaro Watanabe$^{1,6,7,8}$}

\address{$^1$Asia Pacific Center for Theoretical Physics (APCTP), Pohang, Gyeongbuk 37673, Korea}
\address{$^2$Quantum Universe Center, Korea Institute for Advanced Study (KIAS), Seoul 02455, Korea}
\address{$^3$INO-CNR BEC Center and Department of Physics, University of Trento, 38123 Povo, Italy}
\address{$^4$Center for Computational Sciences, University of Tsukuba, Tsukuba 305-8577, Japan}
\address{$^5$RIKEN Nishina Center, Wako, Saitama 351-0198, Japan}
\address{$^6$Department of Physics, POSTECH, Pohang, Gyeongbuk 37673, Korea}
\address{$^7$Center for Theoretical Physics of Complex Systems, Institute for Basic Science (IBS), Daejeon 34051, Korea}
\address{$^8$University of Science and Technology (UST), 217 Gajeong-ro, Yuseong-gu, Daejeon 34113, Korea}
\ead{gentaro@ibs.re.kr}


\vspace{10pt}

\date{\today}
\begin{abstract}
We study multiple period states of a two-component unpolarized superfluid Fermi gas in an optical lattice along the Bardeen-Cooper-Schrieffer (BCS) to Bose-Einstein condensate (BEC) crossover. The existence of states whose period is a multiple of the lattice spacing is a direct consequence of the non-linear behavior of the gas, which is due to the presence of the order parameter associated with superfluidity. By solving Bogoliubov-de Gennes equations for a superfluid flow with finite quasimomentum, we find that, in the BCS side of the crossover, the multiple period states can be energetically favorable compared to the normal Bloch states and their survival time against dynamical instability drastically increases, suggesting that these states can be accessible in current experiments, in sharp contrast to the situation in BECs.
\end{abstract}

%

\section{Introduction}
\label{sec:intro}

Density structures and patterns caused by the interplay of competing effects are ubiquitous in nature. 
Examples are the competition between dispersion and non-linearity, which yields solitons \cite{soliton}, the competition between crystalline order and Peierls instability of conduction electrons, which results in charge and spin density waves \cite{gruner}, or the Fulde-Ferrell-Larkin-Ovchinnikov (FFLO) state \cite{fflo}, with a spatially-dependent pairing field originating from the competition between the mismatch of the Fermi surfaces in the imbalanced systems and the energy gain by the condensation; but similar conditions also occur in the ``pasta'' phases in neutron stars \cite{pasta}, in nuclear halo \cite{halo}, and in superfluid $^4$He \cite{pitaevskii}. In the case of superfluids in a periodic lattice, non-linearity due to the presence of the order parameter, which favors a quadratic energy dispersion, can lead to the persistence of the quadratic-like dispersion beyond the Brillouin zone edge and give rise to non-trivial loop structure called ``swallowtail'' in the energy band \cite{wu_st,diakonov,mueller,machholm,BEC_Kronig,danshita,swallowtail,eckel}.
Due to their high controllability \cite{pethick_smith,lattice}, ultracold gases offer an excellent test bed for exploring these intriguing phenomena.

For atomic Bose-Einstein condensates (BECs) flowing in periodic potentials with finite quasimomentum, it was found that non-linearity of the interaction can cause the appearance of stationary states whose period is not equal to the lattice constant as in the usual Bloch states, but is a multiple of it \cite{machholm04,pethick_smith,maluckov};  such states are called multiple (or $n$-tuple) period states. In BECs without long-range interaction, however, these states are energetically unfavorable compared to the normal Bloch states and unstable against small perturbations \cite{machholm04}. Here we investigate multiple period states in atomic Fermi superfluids, which are particularly interesting for their analogs in condensed matter physics and nuclear physics, such as superconducting electrons in solids and superfluid neutrons in neutron stars \cite{gezerlis,vc_crossover}. Furthermore, by using Feshbach resonances one can continuously go from the Bardeen-Cooper-Schrieffer (BCS) to the BEC regimes \cite{bloch,giorgini}, thus allowing one to understand Bose and Fermi superfluids from a unified perspective. Unlike the case of Bose gases, little has been studied about multiple period states in Fermi gases and their existence itself along the BCS-BEC crossover is an open question. In this work, we show that they indeed exist and they can be energetically favorable compared to the normal Bloch states in the BCS regime. Furthermore, we find that, despite being dynamically unstable, their lifetime becomes drastically long by going toward the deep BCS limit, possibly allowing for their experimental observation.

This paper is organized as follows. After we explain the basic formalism employed in the present work in section \ref{sec:sys}, we show that multiple period states appear in Fermi superfluids along the BCS-BEC crossover and discuss their stationary properties in section \ref{sec:statsol} and \ref{sec:energy}. We then discuss their dynamical stability in section \ref{sec:dyn}. Finally, this paper is concluded in section \ref{sec:conclusion}.

\section{Setup and basic formalism}
\label{sec:sys}

We consider an equally populated (unpolarized) two-component Fermi gas in the superfluid phase at zero temperature, moving in a one-dimensional (1D) optical lattice,
\begin{equation}
  V_{\rm ext}({\bf r}) = V_{\rm ext}(z) \equiv V_0 \sin^2{q_{B}z}  = s E_{R} \sin^2{q_{B}z}\, ,
\end{equation}
where $s$ is the dimensionless parameter of the lattice height, $E_{R}\equiv \hbar^2q_{B}^2/2m$ is the recoil energy, $m$ is the atom mass, $q_{B}\equiv\pi/d$ is the Bragg wave vector, and $d$ is the lattice constant. Note that $q_B$ differs from the fundamental vector of a 1D reciprocal lattice, $2\pi/d$, by a factor of $2$. The gas is uniform in the transverse directions and we look for stationary states of the system in the BCS-BEC crossover by numerically solving the Bogoliubov-de Gennes (BdG) equations \cite{degennes,giorgini}:
\beq 
\left( \begin{array}{cc}
H(\mathbf r) & \Delta (\mathbf r) \\
\Delta^\ast(\mathbf r) & -H(\mathbf r) \end{array} \right)
\left( \begin{array}{c} u_i( \mathbf r) \\ v_i(\mathbf r)
\end{array} \right)
=\epsilon_i\left( \begin{array}{c} u_i(\mathbf r) \\
v_i(\mathbf r) \end{array} \right) \; ,
\label{eq_BdG}
\eeq
where 
$H(\mathbf r) =-\hbar^2 \nabla^2/2m +V_{\rm ext}(\mathbf
r)-\mu$, 
$u_i(\mathbf r)$ and $v_i(\mathbf r)$ are quasiparticle
amplitudes, and $\epsilon_i$ is the corresponding quasiparticle
energy.
The chemical potential $\mu$ is determined from the
constraint on the average density $n_0\equiv N/V = V^{-1} \int n(\mathbf r)\, d{\bf r} = 2V^{-1} \sum_i \int \left| v_i(\mathbf r) \right|^2d{\bf r}$ with $N$ being the number of particles and $V$ being the volume,
and the pairing field $\Delta(\mathbf r)$ should 
satisfy a self-consistency condition $ \Delta(\mathbf r) =
-g \sum_i u_i(\mathbf r) v_i^*(\mathbf r)$,  where $g$ is the coupling 
constant for the $s$-wave contact interaction which 
needs to be renormalized \cite{randeria,bruun,bulgac02,grasso}.
The total energy $E$ is given by
\beq
  E = \int d{\mathbf r} \left[ \frac{\hbar^2}{2m}\left(2\sum_i |\nabla v_i(\mathbf r)|^2\right) + V_{\rm ext}(\mathbf r) n(\mathbf r) + \frac{1}{g} |\Delta(\mathbf r)|^2 \right]\, .\nonumber
\eeq

In this formalism, a stationary motion of the superfluid in the $z$-direction, relative to the infinite periodic potential at rest, is described by solutions of equation (\ref{eq_BdG}) with quasimomentum $P$ per atom (not per pair), or the corresponding wave vector $Q=P/\hbar$, such that the quasiparticle amplitudes can be written in the Bloch form as
$u_i(\mathbf r) =
\tilde{u}_i(z) e^{i Q z}e^{i\mathbf k \cdot \mathbf r }$ and
$v_i(\mathbf r) = \tilde{v}_i(z) e^{-i Q z}e^{i\mathbf k \cdot
\mathbf r }$
leading to the pairing field as 
$\Delta(\mathbf r)=e^{i 2Q z}\tilde{\Delta}(z)$.
Here $\tilde{\Delta}(z)$, 
$\tilde{u}_i(z)$, and $\tilde{v}_i(z)$ 
are complex functions with period $\nu$ times 
$d$, with $\nu\in\{1, 2, 3, \cdots\}$, and
the wave vector $k_z$ lies in the 
first Brillouin zone for a supercell (a cell containing several 
primitive cells) with period $\nu$, i.e., 
$|k_z| \le q_{B}/\nu$. 
This Bloch decomposition transforms 
(\ref{eq_BdG}) into the following BdG equations for $\tilde{u}_i(z)$ 
and $\tilde{v}_i(z)$:
\beq
\left( \begin{array}{cc}
\tilde{H}_{Q}(z) & \tilde{\Delta}(z) \\
\tilde{\Delta}^\ast(z) & -\tilde{H}_{-Q}(z) \end{array} \right)
\left( \begin{array}{c} \tilde{u}_i(z) \\ \tilde{v}_i(z)
\end{array} \right)
=\epsilon_i\left( \begin{array}{c} \tilde{u}_i(z) \\
\tilde{v}_i(z) \end{array} \right) \;,
\label{eq_BdG2}
\eeq
where
\beq
  \tilde{H}_{Q}(z)\equiv \frac{\hbar^2}{2m} \left[ k^2_\perp
+\left(-i\partial_z+Q+k_z\right)^2 \right] +V_{\rm ext}(z) -\mu\, .
\nonumber\label{hq}
\eeq
Here, $k_\perp^2\equiv k_x^2 + k_y^2$ and
the label $i$ represents the wave vector $\mathbf k$ as well 
as the band index. We solve this BdG equations (\ref{eq_BdG2}) for a 
supercell with period $\nu$ ($-\nu d/2 \le z \le \nu d/2$) to obtain the period-$\nu$ states. As in \cite{optlatunit,vc,vc_crossover,swallowtail} (see also appendix~iv), the detailed procedure follows these steps: Starting from an initial guess of $\tilde{\Delta}(z)$ and $\mu$ (final results are robust to the choice of the initial guess), we diagonalize the matrix of the left-hand side of the BdG equations (\ref{eq_BdG2}) and obtain $\epsilon_i$, $\tilde{u}_i$, and $\tilde{v}_i$. Based on the obtained $\tilde{u}_i$ and $\tilde{v}_i$, we calculate the average number density $n_0$ and $\tilde{\Delta}(z)$. If the resulting $n_0$ does not agree with a given target value $n_0^{\rm target}$, we update $\mu$ according to the difference between these values. Specifically, in our calculations, we set the updated value $\mu'$ using the following formula: $\mu' = \mu\ (n_0^{\rm target}/n_0)^{\eta}$ with $\eta < 1$ such as $\eta = 2/3$, $1/3$, $1/5$, etc. Until $\mu$ converges within $\sim 10^{-7} E_R$, we iterate the above procedure using the obtained $\tilde{\Delta}(z)$ and updated $\mu$.
We check that other key quantities also converge with sufficient accuracy.

In the following, we mainly present the results for $s = 1$ and $2$ 
with $E_F/E_R =0.25$ as examples, where $E_{F} = \hbar^2k_{F}^2/(2m)$ 
and $k_{F}=(3\pi^2 n_0)^{1/3}$ are the Fermi energy and 
wavenumber of a uniform free Fermi gas of density $n_0$. These values 
fall in the range of parameters of feasible experiments \cite{miller}. 
We have performed systematic calculations for different values of 
$E_F/E_R$ (0.2, 0.25, and 0.5) as well and checked that the main results 
of stationary properties remain qualitatively the same. We denote by $P_{\rm edge}$ the 
quasimomentum $P$ per atom at the edge of the Brillouin zone for the normal 
Bloch states with period $1$. 
For superfluids of fermionic atoms, $P_{\rm edge}\equiv\hbar q_{B}/2$. Note that this value differs by a factor of $2$ from that of superfluids of bosonic atoms, $\hbar q_{B}$, because the elementary constituents of Fermi superfluids are pairs of fermionic atoms.

\section{Stationary solutions}
\label{sec:statsol}

\begin{figure}[!tb]
\centering
\rotatebox{270}{
\resizebox{!}{16cm}
{\includegraphics{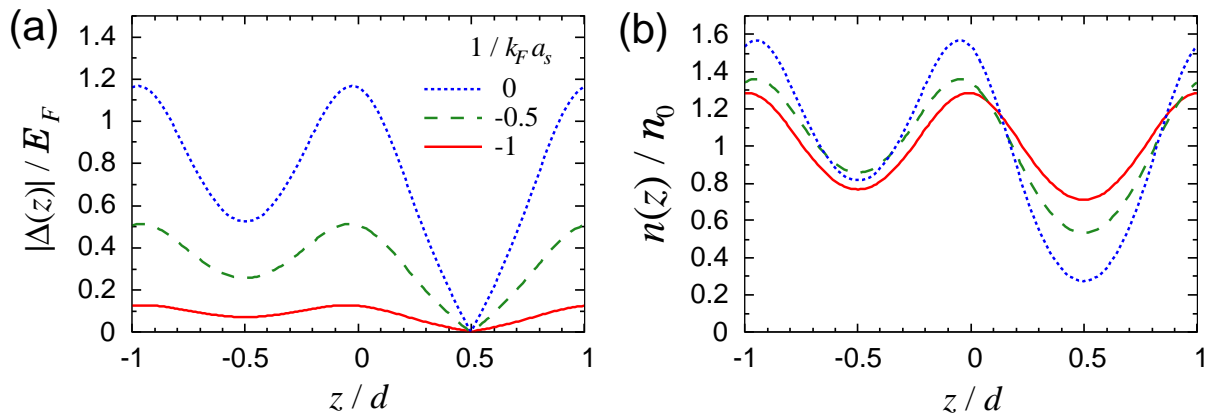}}}
\caption{\label{fig_profiles}
  Profiles of (a) the magnitude of the 
  pairing field $|\Delta(z)|$ and (b) the density $n(z)$ of the lowest
  period-doubled states with period $2d$ (period-2 states) 
  at their Brillouin zone edge $P=P_{\rm edge}/2=\hbar q_{B}/4$ 
  for three different values of $1/k_{F}a_s$:
  $1/k_{F}a_s =-1$ (red solid line), $-0.5$ (green dashed line), and
  $0$ (blue dotted line).  They are obtained for $s=1$ and
  $E_{F}/E_{R}=0.25$.  At $P=P_{\rm edge}/2$, $\Delta(z)$ of the
  period-2 states has a node. 
}
\end{figure}

\begin{figure}[tb]
\begin{center}
\rotatebox{0}{
\resizebox{11cm}{!}
{\includegraphics{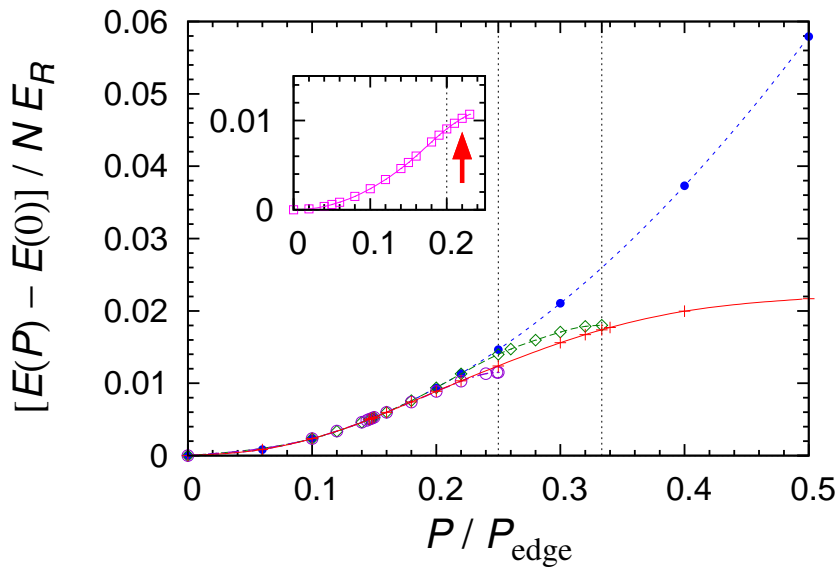}}}
\caption{\label{fig_e}
Energy $E$ per particle in units of $E_R$
as a function of the quasimomentum $P$.
Here we set $s=1$, $E_F/E_R =0.25$,
and $1/k_F a_s =-1$.
The normal Bloch states with period $d$ are shown by the blue dotted line with $\bullet$ symbols, and the multiple period states are shown by the red solid line with $+$ (period 2), the green dashed line with {\Large$\diamond$} (period 3), the purple dashed-dotted line with {\Large$\circ$} (period 4), and the magenta solid line with $\square$ (period 5).
The dotted lines at $P/P_{\rm edge}=1/3$, $0.25$, and $0.2$ show the first Brillouin zone edge for the period-3, 4, and 5 states, respectively.
The inset shows that the lowest band of the period-5 states continues beyond the first Brillouin zone edge and the swallowtail appears (shown by the red arrow).
}
\end{center}
\end{figure}

In figures~\ref{fig_profiles}(a) and \ref{fig_profiles}(b), we show 
the profiles of the pairing field $|\Delta(z)|$ and the number density $n(z)$
of the lowest period-doubled states (period-2 states), respectively.
Here we set $P=P_{\rm edge}/2=\hbar q_B/4$ at the Brillouin zone edge
of the period-2 states, where the feature of the period-2 states appears most prominently \footnote{For the BEC case it has been shown that, with increasing non-linearity (i.e., the interaction strength $g_b$) from the linear limit ($g_b=0$), the period-doubled states start to appear at the Brillouin zone edge of the period-2 system and their band extends in the Brillouin zone \cite{machholm04}. In this sense, the Brillouin zone edge for the period-2 system is a representative point for period-doubled states.}.
The feature of the period doubling and the difference between 
the regions of 
$-1<z/d\le0$ and $0<z/d\le1$ can be clearly seen in $|\Delta(z)|$
at any value of $1/k_F a_s$.
At $P=P_{\rm edge}/2$, $\Delta(z)$ of the period-2 states has
a node [see $z/d=0.5$ in figure~\ref{fig_profiles}(a)] 
and consequently the supercurrent is zero
($\partial_PE=0$; see figure~\ref{fig_e}). 
On the other hand, especially in the deeper BCS side ($1/k_F a_s =-1$), the difference in $n(z)$ between the regions of $-1<z/d\le0$ and $0<z/d\le1$ is small [see the red line in figure~\ref{fig_profiles}(b)]. Around $z/d=0.5$, where $|\Delta(z)|$ vanishes, the density remains large, suggesting the existence of Andreev-like localized states. The density difference between the two regions becomes larger with increasing $1/k_F a_s$ toward the BEC regime.

\begin{figure}[t]
\centering
\rotatebox{270}{
\resizebox{!}{16cm}
{\includegraphics{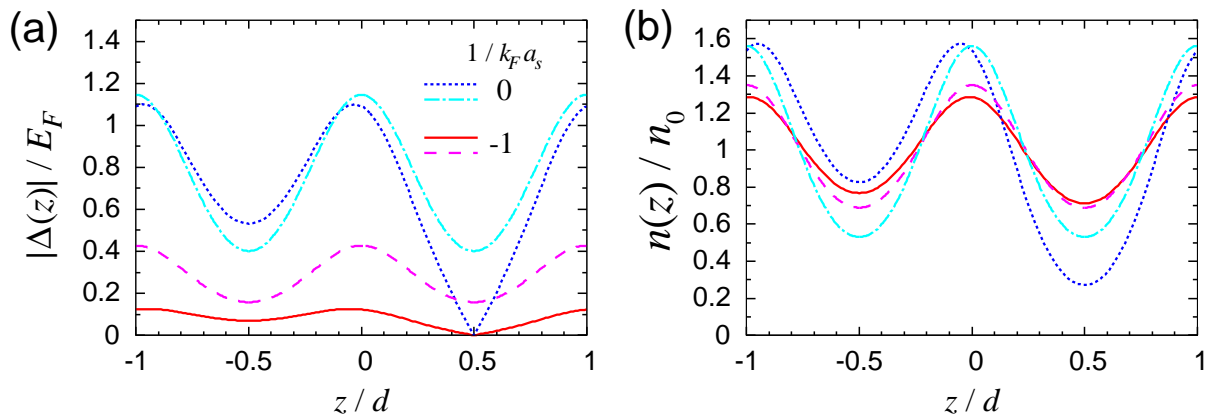}}}
\caption{Profiles of (a) the pairing field $|\Delta(z)|$ and (b) the density $n(z)$ of the normal Bloch states (magenta dashed line for $1/k_Fa_s=-1$ and cyan dashed-dotted line for $1/k_Fa_s=0$) and period-doubled states (red solid line for $1/k_Fa_s=-1$ and blue dotted line for $1/k_Fa_s=0$) at $P=P_{\rm edge}/2$. Here we set $s=1$ and $E_F/E_R=0.25$ as in figure~\ref{fig_profiles}.
}
\label{fig_profiles_supp}
\end{figure}

In figure~\ref{fig_profiles_supp}, we compare the profiles of $|\Delta(z)|$ and $n(z)$ between the normal Bloch states and the period-doubled states for $1/k_Fa_s=-1$ and $0$. Since we set $P$ at the Brillouin zone edge of the period-doubled states ($P=P_{\rm edge}/2$), where their supercurrent is zero, $\Delta(z)$'s of the period-doubled states have a node while those of the normal Bloch states do not. Note that, in the BCS regime of $1/k_Fa_s=-1$, $n(z)$'s of the normal Bloch state and the period-doubled state are almost the same, but $|\Delta(z)|$'s are significantly different.

It is instructive to consider the deep BCS limit of $1/k_Fa_s \rightarrow -\infty$. There, $\Delta(z)$, which is the origin of the non-linearity, vanishes and $n(z)$ of the neighboring sites becomes identical so that the nature of the period doubling disappears in this limit. We observe that by going to the deep BCS regime, where $\Delta(z)$ and the supercurrent are infinitesimally small, the energy difference $E(P)-E(0)$ for period-2 states decreases [i.e., $E(P)$ becomes more flat] and period-2 states at $P=P_{\rm edge}/2$ approach the normal Bloch state at $P=0$.
These observations are consistent with the fact that, if $\Delta(z)=0$, our non-linear BdG equations reduce to the linear Schr\"odinger equation, whose solutions have the periodicity of the lattice due to the Bloch theorem. Multiple period states are hence possible only in the superfluid phase. 
It is worth mentioning here that these multiple-period states are essentially different from the FFLO \cite{fflo} or soliton lattice \cite{solitonlat} states in the imbalanced (polarized) systems. In our case, the non-trivial spatial dependence of the pairing field is a purely non-linear phenomenon caused by the presence of the superfluid order parameter, while in the other cases it is due to the non-zero center-of-mass momentum of the pair, which requires the mismatch of the Fermi surfaces between two components. The multiple period states studied in the present work is also different from the charge density wave due to the nesting of the Fermi surface (see appendix~i for details).

As a final comment on the spatial structure of stationary solutions, it is worth noting that the presence of a node in $\Delta$ is a sufficient condition for a zero supercurrent, but it is not a necessary condition. For example, at the Brillouin zone center, the supercurrent is of course zero because the phase of $\Delta$ is constant ($P=0$), but $\Delta$ does not have a node. On the other hand, for nonzero $P$, the phase of $\Delta$ depends on the position. Therefore, when the supercurrent is zero at the Brillouin zone boundary, $\Delta$ must have a node. States with more nodes in $\Delta$ have higher energy in general. As to the lowest periodic states which we discuss in the present work, the number of nodes is thus one per supercell at the Brillouin zone boundary. To minimize the energy, the node is located at the potential maximum.

\section{Energetics}
\label{sec:energy}

In figure~\ref{fig_e}, we plot the energy per particle of the lowest band as a function of the quasimomentum $P$ for the normal Bloch states (blue dotted line) and the multiple period states. 
We show the results at $1/k_F a_s =-1$ in the BCS side. 
In the region of small $P$, all lines of the multiple period states collapse onto the line of the normal Bloch states, as they are all equivalent in this region, the states with period 1 being just a subset of all the multiple period states \footnote{The critical quasimomentum $P_c$ for the pair-breaking instability of the normal Bloch state is $0.147 P_{\rm edge}$ in this case. Note that $P_c$ seems to coincide with the value at which the multiple period states start to separate from the normal Bloch state. This suggests that emergence of the multiple period states lead to the Landau instability of the normal Bloch state.}. 

Conversely, the multiple period states for small $\nu$ ($\nu\le 4$ in the case of figure~\ref{fig_e}) become energetically more stable than the normal Bloch states near the first Brillouin zone edge of each multiple period state, i.e., $P \lesssim P_{\rm edge}/\nu$ for period-$\nu$ states. In particular, the period-doubled states are the lowest in energy in a wide range of $P$ [In the case of figure~\ref{fig_e}, period-2 states are always energetically lower than period-3 states, which holds in the region of $0.245\lesssim E_F/E_R \lesssim 0.4$ for the same values of $s=1$ and $1/k_Fa_s=-1$ (see appendix~ii for details).]. 
This is in striking contrast to the situation in BECs and in the BEC regime of the BCS-BEC crossover (see later), where the lowest band of normal Bloch states is always lower in energy than the multiple period states \cite{machholm04}; the latter appear as an upper branch of the swallowtail band structure (with $\partial_P^2 E >0$) around the Brillouin zone edge of the respective multiple period states \cite{machholm04}. Figure~\ref{fig_e} shows instead that, in the BCS regime, the lowest band of the multiple period states for small $\nu$ has $\partial_P^2 E<0$ near the Brillouin zone edge $P\lesssim P_{\rm edge}/\nu$.

At first sight, this seems to imply a pathological situation in which $\nu$-period states with large $\nu$ would be lower in energy than the normal Bloch states even in the limit of $P\rightarrow 0$.
However, this pathological situation is saved by the emergence of the swallowtail: The multiple period states with large $\nu$ continue being almost identical to the normal Bloch states, and keep their nearly quadratic dispersion around $P \sim 0$ even beyond their first Brillouin zone edge at $P_{\rm edge}/\nu$, which results in the swallowtail band structure for the period-$\nu$ system. In the case of figure~\ref{fig_e}, the swallowtail starts to appear at $\nu=5$ (see the inset of figure~\ref{fig_e}).

\begin{figure}[!tb]
\begin{center}
\rotatebox{0}{
\resizebox{11cm}{!}
{\includegraphics{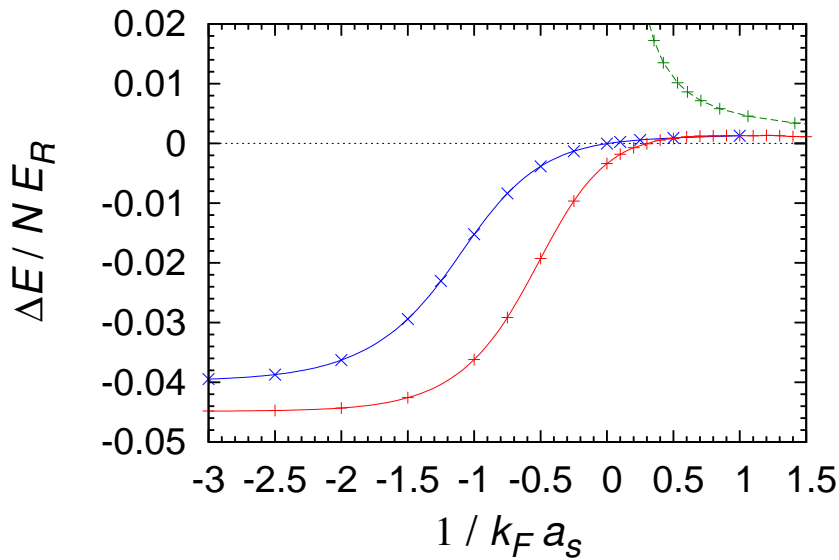}}}
\caption{\label{fig_de}
Difference $\Delta E \equiv E_2-E_1$ of the total energy per particle in units of $E_{R}$
between the normal Bloch states ($E_1$) and period-doubled states ($E_2$) at $P=P_{\rm edge}/2$.
The parameters we have used are $s=1$, $2$ and $E_{F}/E_{R}=0.25$.
The red solid line with $+$ ($s=1$) and blue solid line with $\times$ ($s=2$) show the results obtained by solving the BdG equations
and the green dashed line shows the results by the Gross-Pitaevskii equations 
for parameters corresponding to $s=1$ and $E_F/E_R=0.25$.
}
\end{center}
\end{figure}

In figure~\ref{fig_de}, we show the total energy difference $\Delta E\equiv E_2-E_1$ between the normal Bloch states ($E_1$) and the period-2 states ($E_2$) at $P=P_{\rm edge}/2=\hbar q_{B}/4$ along the BCS-BEC crossover. As we have seen in figure~\ref{fig_e}, the period-doubled states are energetically more stable (i.e., $\Delta E<0$) in the BCS regime. With $1/k_{F}a_s$ increasing from the deep BCS limit, $\Delta E$ increases from a negative value and finally period-doubled states become higher in energy than normal Bloch states (i.e., $\Delta E>0$) in the BEC side. 
Here we point out that, in the region of $\Delta E<0$, the period-doubled states form a band which is convex upward and smoothly connects to that of the normal Bloch states (see figure \ref{fig_e}), and the swallowtail does not exist. On the other hand, in the region of $\Delta E>0$, they form a concave upper edge of the swallowtail, which is located above the crossing point (``$\times$''-like shape) of the swallowtail. Therefore, hysteresis caused by the swallowtail, which could be observed in the latter region ($\Delta E>0$) of the BCS-BEC crossover and in BECs \cite{eckel}, would disappear in the former region ($\Delta E<0$).

We also show the results in the BEC side obtained by solving the Gross-Pitaevskii (GP) equation for corresponding parameters (the green dashed line).
Namely, GP equation for bosons of mass $m_b=2m$ interacting with scattering length $a_b\equiv \alpha a_s =2a_s$ for the mean-field theory in an optical lattice $2V_{\rm ext}(z)$. We can relate $1/k_{F}a_s$ and $g_bn_b/E_{{R}, b}$, where $g_b\equiv 4\pi\hbar^2a_b/m_b$, $n_b=n/2$, and $E_{{R}, b}\equiv \hbar^2(\pi/d)^2/2m_b$ are the interaction parameter, the average density, and the recoil energy of bosons, as $1/k_{F}a_s = (4\alpha/3\pi) \left(g_bn_b/E_{{R}, b}\right)^{-1} \left(E_F/E_R\right)$.
We note that $\Delta E$ of the GP results approaches zero 
from positive values with increasing $1/k_{F}a_s$.
This suggests that, before the BdG results (the red and blue solid lines) 
converge to the GP ones in the deep BEC regime, $\Delta E$ takes a maximum value.

For different strength $s$ of the lattice, we see that the curve of $\Delta E$ is somewhat shifted towards the BCS side with increasing $s$, so the period-doubled states become less stable (see the red solid line with $+$ for $s=1$ and the blue solid line with $\times$ for $s=2$ in figure \ref{fig_de}).
This might be due to the formation of bosonic molecules of fermionic atoms induced by the external lattice potential \cite{fedichev04,orso05,optlatunit}.

The energetic stability of multiple period states in the BCS regime can be physically understood as follows. Let us consider the different behavior of $\Delta(z)$ and $n(z)$ for a period-2 state and a normal Bloch state at $P=P_{\rm edge}/2$. In the case of a normal Bloch state, since $|\Delta(z)|$ is exponentially small in the BCS regime, we can distort the order parameter $\Delta(z)$ to produce a node, like the one in the period-2 state, with a small energy cost (per particle) up to the condensation energy $|E_{\rm cond}|/N \ll E_F$, where $E_{\rm cond} \equiv g^{-1} \int d^3r\, |\Delta({\bf r})|^2$. However, making a node in $\Delta(z)$ kills the supercurrent $j=V^{-1}\partial_P E$, which yields a large gain of kinetic energy (per particle) of the superfluid flow of order $\sim P_{\rm edge}^2/m \sim E_R$. Even if $\Delta(z)$ is distorted substantially to have a node, the original density distribution of the normal Bloch state is almost intact so that the increase of the kinetic energy and the potential energy due to the density variation is small. Therefore, the period-2 state is energetically more stable than the normal Bloch state in the BCS regime.
In the above discussion, the key point is that $\Delta(z)$ and $n(z)$ can behave in a different way in the BCS regime.
On the other hand, in the BEC limit, the density is directly connected to the order parameter as $n(z)\propto |\Delta(z)|^2$, and distorting the order parameter accompanies an increase of the kinetic and potential energies due to a large density variation.

More generally, for period-$\nu$ states in comparison with a normal Bloch state at $P=P_{\rm edge}/\nu$, the energy cost to distort $\Delta(z)$ to have a node is up to $\sim |E_{\rm cond}|/N \ll E_F$, but the energy gain is of order $\sim P_{\rm edge}^2/m\nu^2 \sim E_R/\nu^2$, which is reduced by a factor of $\nu^2$. We thus see that period-$\nu$ states with sufficiently large $\nu$ cannot be energetically more stable than the normal Bloch state as has been observed before.

\section{Dynamical stability and survival time}
\label{sec:dyn}

So far, we have seen that multiple-period states exist as energetically stable stationary solutions of the BdG equations. The next important issue is their dynamical stability, that is, whether and how long they can survive under small perturbations, which are unavoidable in experiments. We face this problem by performing numerical simulations based on the time-dependent BdG (TD-BdG) equations. 

A crucial difference between the stationary calculations (time-independent BdG) in the previous sections and the dynamical (time-dependent BdG) calculations in this section is the following. The ideal configuration to study the stationary solutions with a given periodicity $\nu$ is a supercell with $\nu$ sites under the Bloch-wave boundary conditions, as we have done in the previous sections. Conversely, dynamical calculations has to account for excited states with any wavelength, possibly including long wavelength perturbations which may trigger a dynamical instability, so that the Bloch-wave boundary conditions cannot be used. We instead solve the TD-BdG equations in a large computational box of length $L_z$ in the $z$-direction, including a sufficiently large number of supercells of $\nu$ sites, with the periodic boundary conditions, in order to mimic an infinite system with good enough accuracy. We use $L_z = 32d$ to $64d$; $L_z$ is chosen to be a multiple of $8d$ for convenience, so that the allowed values of the wave vector $k_z$ discretized as $\Delta k_z = 2\pi/L_z$ are commensurate with the value of the quasimomentum at the first Brillouin zone edge for the period-2 states, $P_{\rm edge}/(2\pi) = q_B/4$. Finally, stationary BdG calculations for such large boxes with the periodic boundary conditions are not feasible with our current computational resources, since they require long iterative procedures for many values of $P$; a direct comparison between stationary and time-dependent BdG results would be possible only for smaller values of $L_z$, corresponding to less than about ten lattice sites, for which the extrapolation to an infinite system would be unreliable.

As initial configuration of the TD-BdG simulations, we use a configuration based on the stationary solution of the BdG equations, which is constructed as follows. Among the quasiparticle amplitudes $u_i$ and $v_i$ obtained by solving equation~(\ref{eq_BdG2}), we select those with (quasi)wave vectors $k_z$ equal to multiples of $\Delta k_z = 2\pi/L_z$. In this way, we construct the approximate stationary solution of the BdG equations~(\ref{eq_BdG}) with the periodic boundary condition. 
Then we integrate the TD-BdG equations using a $4$-th order predictor-corrector method. The basic structure of the code is the same as the one in \cite{code}.

\begin{figure}[!tb]
\rotatebox{0}{
\resizebox{!}{6cm}
{\includegraphics{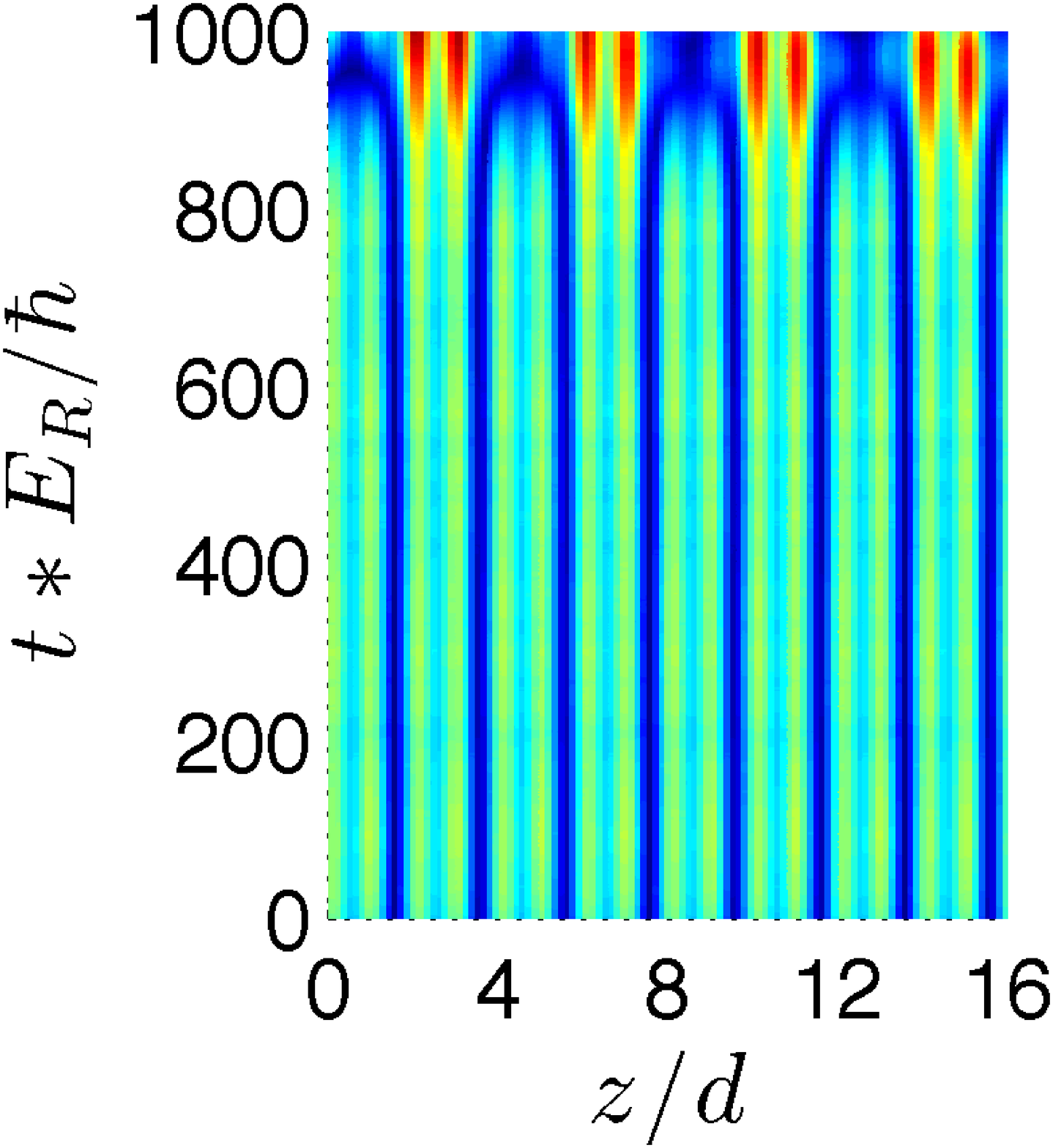}}}
\rotatebox{0}{
\resizebox{!}{6cm}
{\includegraphics{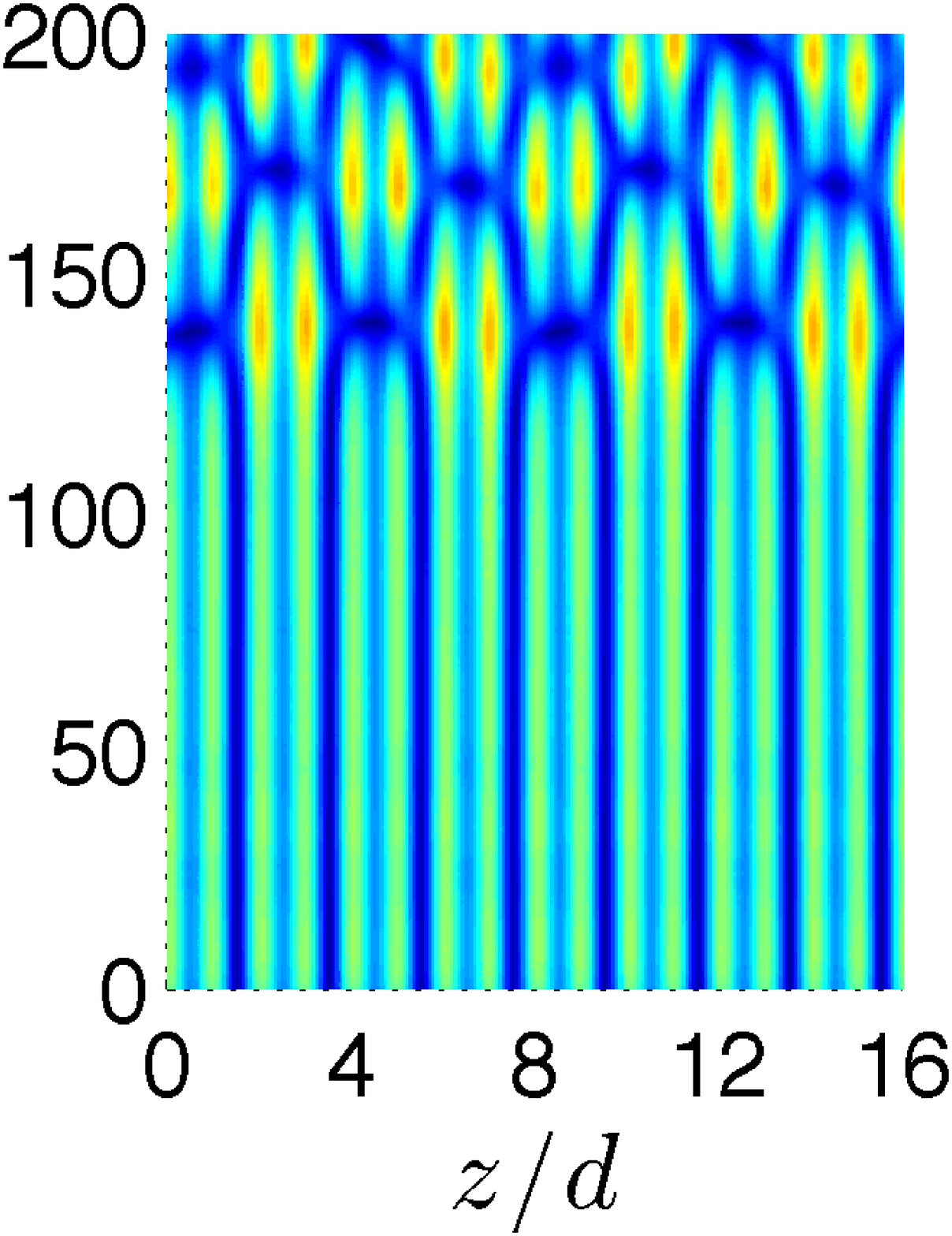}}}
\rotatebox{0}{
\resizebox{!}{6cm}
{\includegraphics{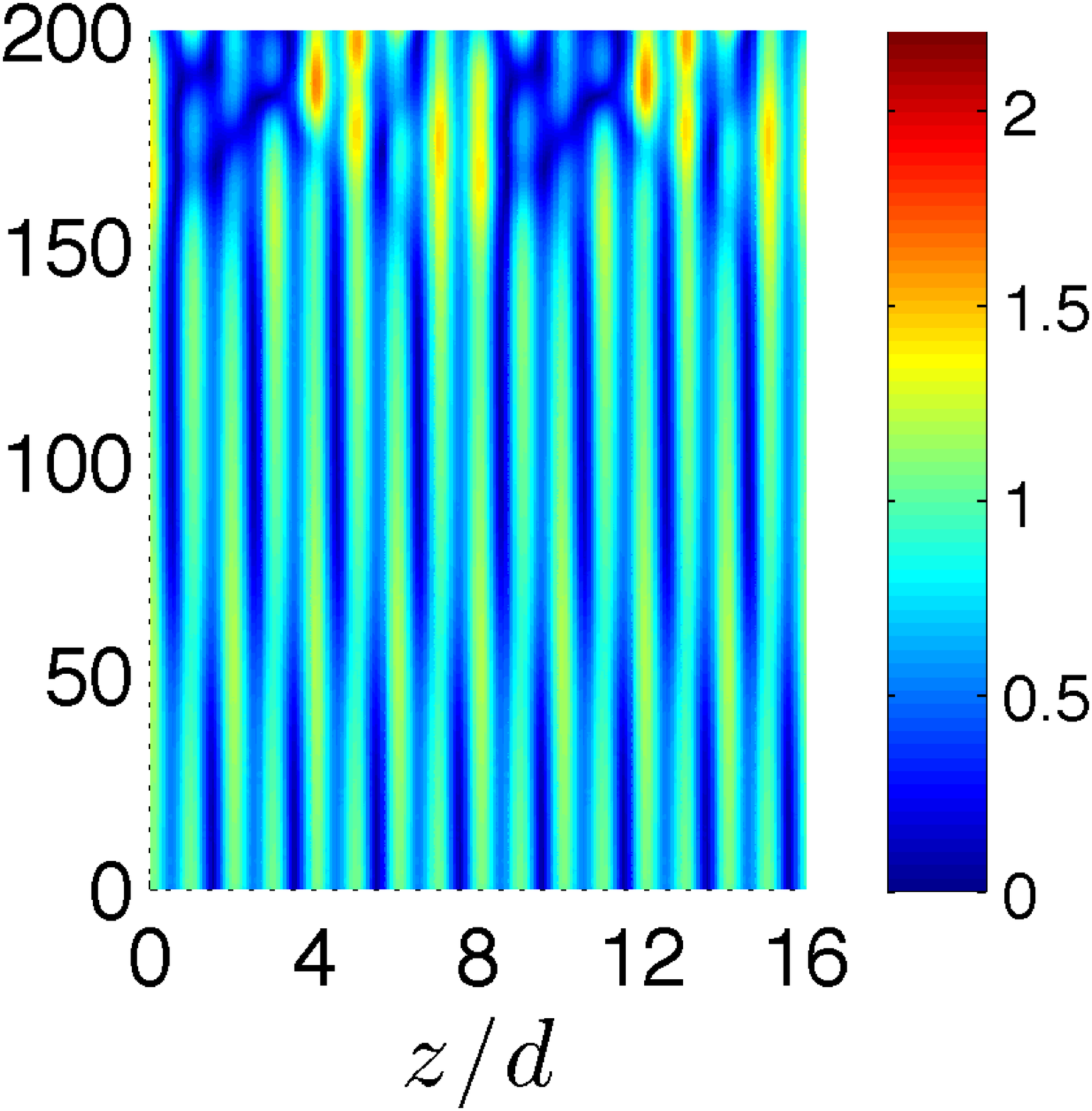}}}
\caption{\label{fig_evolution}
Time evolution of the magnitude of the pairing field $|\Delta(z,t)|/|\Delta(z=0,t=0)|$ of the period-doubled state at $P=P_{\rm edge}/2$ for $1/k_Fa_s=-1$ (left panel), $0$ (middle panel), and $0.5$ (right panel). Here, $s=1$ and $E_F/E_R=0.25$.
Actual calculation has been done for $L_z=32d$ in the cases of $1/k_Fa_s=0.5$ and $0$ and for $L_z=48d$ in the case of $1/k_Fa_s=-1$; a part of the system is shown in the figure.
}
\end{figure}

\begin{figure}[!tb]
\begin{center}
\rotatebox{0}{
\resizebox{13cm}{!}
{\includegraphics{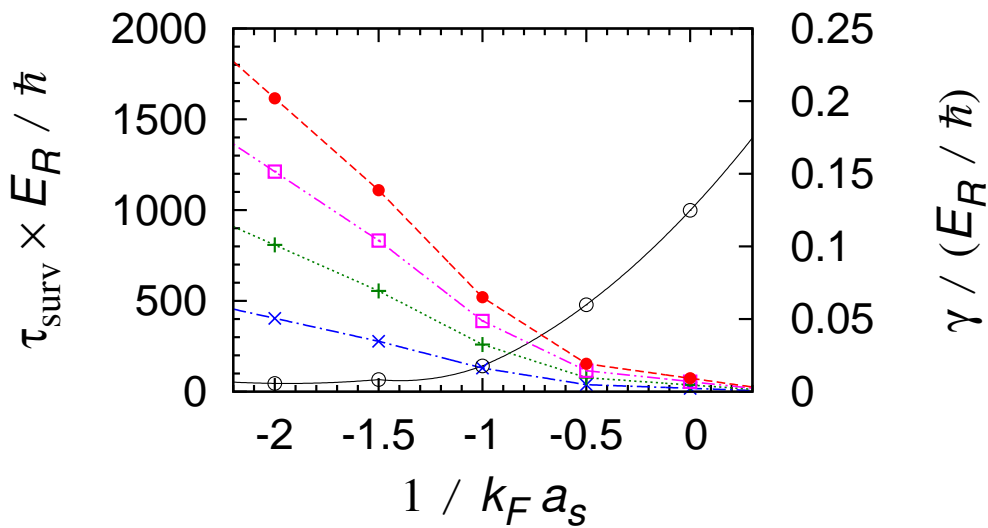}}}
\caption{\label{fig_survival}
Growth rate $\gamma$ of the fastest growing mode (black solid line) and survival time $\tau_{\rm surv}$ of the period-doubled state at $P=P_{\rm edge}/2$ ($s=1$ and $E_F/E_R=0.25$). Blue dashed-dotted, green dotted, magenta dashed double-dotted, and red dashed lines show $\tau_{\rm surv}$ for relative amplitude $\tilde{\eta}(0)$ of the initial perturbation of $10\%$, $1\%$, $0.1\%$, and $0.01\%$, respectively. (The regions of $1/k_Fa_s <-2$ and $>0$ are extrapolated.)
}
\end{center}
\end{figure}

The right and middle panels of figure~\ref{fig_evolution} show the time evolution of $|\Delta(z)|$ of the period-doubled state at $P=P_{\rm edge}/2$ in the BEC side ($1/k_Fa_s=0.5$) and at unitarity, respectively. We see that $|\Delta(z)|$ [and $n(z)$] does not keep its initial profile and large-amplitude oscillations triggered by the dynamical instability set on at $t\simeq 55 \hbar/E_R$ in the right panel and $t\simeq 130 \hbar/E_R$ in the middle panel. We also notice that the TD-BdG simulations allow us to identify the spontaneously growing excitations which trigger the instability. The wavelength of the growing mode is $4d$, $4d$, and $2d$ in the case of $1/k_Fa_s=-1$ (left panel of figure~\ref{fig_evolution}), $0$ (middle panel), and $0.5$ (right panel), respectively.

It is remarkable that the survival time $\tau_{\rm surv}$ of the period-doubled states until they are destroyed by the large-amplitude oscillations drastically increases as going toward the BCS side. In the left panel of figure~\ref{fig_evolution}, we show the time evolution of $|\Delta(z)|$ of the period-doubled state at $P=P_{\rm edge}/2$ for $1/k_Fa_s=-1$. 
In this realization, the period-doubled state almost keeps its initial profile of $|\Delta(z)|$ until $t \simeq 900 \hbar/E_R$, and even longer for $n(z)$ because only a small fraction of particles participate in the pairing in the BCS regime.

To further analyze the time scale of the deviation $|\Delta(z,t)|-|\Delta_0(z)|$ from the true stationary state $\Delta_0(z)$, we take its spatial Fourier transform and look for modes with exponentially growing amplitudes $|\eta(t)|=|\eta(0)|\, e^{\gamma t}$. From a fit we extract the growth rate $\gamma$ of the fastest growing mode. The growth rate $\gamma$ corresponds to the imaginary part of the complex eigenvalue for the fastest growing mode obtained by the linear stability analysis \cite{ring,pethick_smith}. This is intrinsic property of the initial stationary state independent of the magnitude of the perturbation.

The resulting $\gamma$ is shown by the black solid line in figure~\ref{fig_survival}, which clearly shows the suppression of $\gamma$ with decreasing $1/k_Fa_s$. In practice, the survival time $\tau_{\rm surv}$ of the period-doubled states depends on the accuracy of their initial preparation. We estimate $\tau_{\rm surv}$ with $\tilde{\eta}(0) e^{\gamma t} \sim 1$, where $\tilde{\eta}(0)$ is the relative amplitude of the initial perturbation with respect to $|\Delta_0|$ for the fastest growing mode. In figure~\ref{fig_survival}, we show $\tau_{\rm surv}$ for four values of $\tilde{\eta}(0)$. This result suggests that if the initial stationary state is prepared within an accuracy of 10\% or smaller, this state safely sustains for time scales of the order of $100\hbar/E_R$ or more in the BCS side,  corresponding to $\tau_{\rm surv}$ of more than the order of a few milliseconds for typical experimental parameters \cite{miller}: For $E_{R, b}=2\pi\times 7.3 \mathrm{kHz}\times \hbar$ used in the experiment of \cite{miller}, $1 \hbar/E_R = 0.0109$ msec. In the deep BCS regime ($1/k_Fa_s \ll -1$), $\tau_{\rm surv}$ increases further and may become larger than the time scale of the experiments, so that the period-doubled states can be regarded as long-lived states and, in addition, since they have lower energy than the usual Bloch states in a finite range of quasimomenta, they could be realized by, e.g., quasi-adiabatically increasing $P$ from the ground state at $P=0$, which is the normal Bloch state.

Finally, it is worth noting that the BCS transition temperature $T_c$ is roughly estimated as $T_c \sim T_F e^{\pi/2k_Fa_s}$ with $T_F \equiv E_F/k_B$ and $k_B$ is the Boltzmann constant. For the above value of $E_R$ used in the experiment of \cite{miller}, $T_c \sim 200$nK at $1/k_Fa_s =-1$, $50$nK at $1/k_Fa_s =-2$, and $10$nK at $1/k_Fa_s =-3$. Therefore, superfluidity can be realized in the whole region shown in figure \ref{fig_survival} in the current experiments.

\section{Conclusion}
\label{sec:conclusion}

We have studied multiple period states, especially period-doubled states, of superfluid Fermi gases in an optical lattice and have found that they can be energetically more stable than the normal Bloch states and their survival time can be drastically enhanced in the BCS side. The multiple period nature distinctly appears in the pairing field, which could be observed by the fast magnetic field sweep technique \cite{regal,horikoshi}. It is also interesting to point out that the emergence of the period-doubled states in the BCS side is closely connected to the disappearance of the swallowtails, which exist in the BEC side (section \ref{sec:energy}). As a consequence, hysteresis of the superfluid circuits \cite{eckel}, which could be observed experimentally in the BEC side, would disappear by sweeping to the BCS regime. We hope our work will stimulate future experimental studies.

\ack
We acknowledge P. Fulde, M. Horikoshi, M. Modugno, and C.~J. Pethick for helpful discussions.
This work was supported by the Max Planck Society, 
MEST of Korea,
Gyeongsangbuk-Do, Pohang City, for the support of 
the JRG at APCTP, by Basic Science Research Program through 
NRF by MEST (Grant No. 2012R1A1A2008028), by Project Code (IBS-R024-D1), and 
by JSPS KAKENHI (Grant No. 24105006). 
FD acknowledges the support of ERC, through the QGBE grant, and of the Provincia Autonoma di Trento.
Part of calculations were performed on RICC in RIKEN.


\section*{Appendix}

In this appendix we provide additional information about the momentum distribution, comparison of the energy between period-2 and -3 states, comments on rational and irrational number periods, and detailed information of the numerical calculations.

\bigskip
\noindent{\bf i) Momentum distribution}

\begin{figure}[h]
\centering
\resizebox{8.2cm}{!}
{\includegraphics{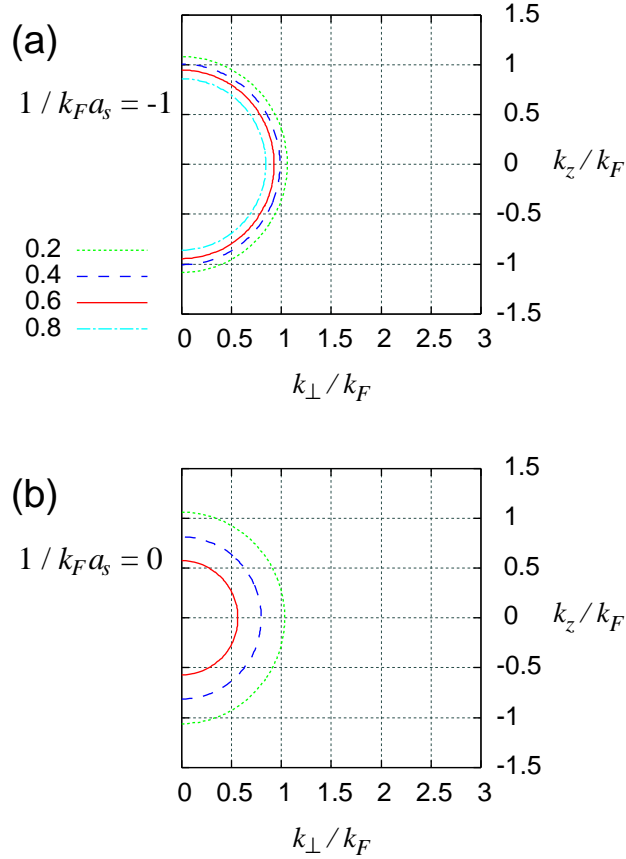}}
\caption{Momentum distribution $n({\bf k}) k_F^3$ at $P=0$ in the BCS side of $1/k_{\rm F}a_s=-1$ [panel (a)] and at unitarity $1/k_{\rm F}a_s=0$ [panel (b)] for $s=1$ and $E_F/E_R=0.25$. The vertical axis shows the wave vector $k_z$ in $z$-direction and the horizontal axis shows the wave vector $k_\perp$ in the transverse directions. Contours at $n({\bf k}) k_F^3 =0.2$ (green dashed-dotted), $0.4$ (blue dashed), $0.6$ (red solid), and 0.8 (cyan dashed) are shown.
}
\label{fig_nqs1_supp}
\end{figure}

\begin{figure}[h]
\centering
\resizebox{8.2cm}{!}
{\includegraphics{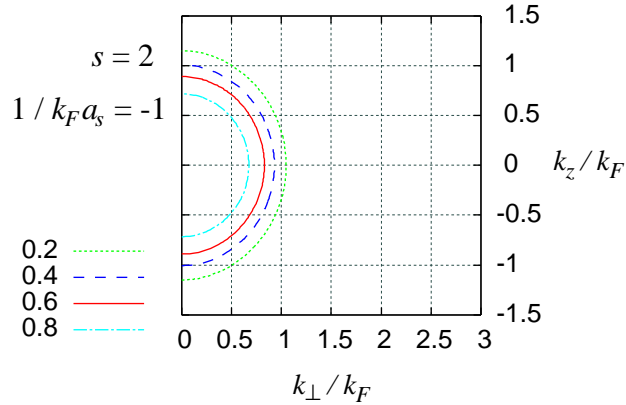}}
\caption{Same as figure~\ref{fig_nqs1_supp} for a stronger periodic potential with $s=2$. As in figure~\ref{fig_nqs1_supp}(a), we set $P=0$, $1/k_{\rm F}a_s=-1$, and $E_F/E_R=0.25$.
}
\label{fig_nqs2_supp}
\end{figure}

The period $n$-tupling studied in the present work is different from the charge density wave due to the nesting of the Fermi surface.
In figure~\ref{fig_nqs1_supp}, we show the momentum distribution $n({\bf k})$ of the normal Bloch states at $P=0$. At $1/k_Fa_s=-1$ [panel (a)], there is a plateau around $k_{\perp}=k_z=0$ with a smeared Fermi surface whose width is characterized by $\sim |\Delta|^{1/2}$, while, at $1/k_Fa_s=0$ [panel (b)], $n({\bf k})$ shows a peak at $k_{\perp}=k_z=0$ rather than a plateau. Note that even though the system is in a periodic potential with nonzero $s$, the Fermi surface is almost spherical.

Figure~\ref{fig_nqs2_supp} is the same as figure~\ref{fig_nqs1_supp}(a), but for a stronger periodic potential with $s=2$. The plateau region of the momentum distribution $n({\bf k})$ is significantly smaller compared to that for $s=1$ shown in figure~\ref{fig_nqs1_supp}(a). This is due to the formation of bound bosonic dimers induced by the stronger periodic potential \cite{fedichev04,orso05,optlatunit}. However, also in this case, $n({\bf k})$ is almost isotropic although it is more compressed in the $k_\perp$-directions compared to the case of $s=1$ [figure~\ref{fig_nqs1_supp}(a)].

\bigskip
\noindent{\bf ii) Comparison of the energy between period-2 and -3 states}

\begin{figure}[ht]
\centering
\resizebox{11cm}{!}
{\includegraphics{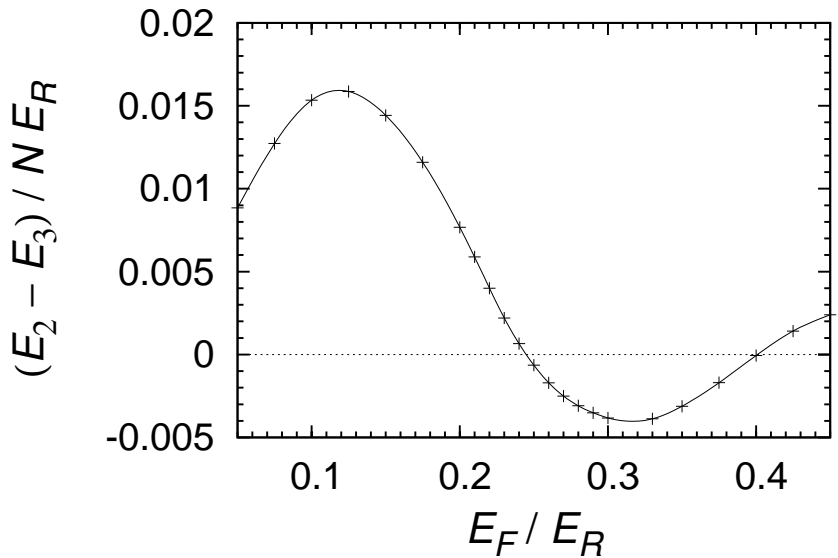}}
\caption{
Energy difference $E_2-E_3$ between the period-2 and -3 states at $P=P_{\rm edge}/3$, where $E_\nu$ ($\nu=2, 3$) represents the energy of the period-$\nu$ state. Here we set $s=1$ and $1/k_Fa_s=-1$, which are the same as in figure~\ref{fig_e}.
}
\label{fig_de_per2and3_supp}
\end{figure}

In figure~\ref{fig_de_per2and3_supp}, we show the energy difference $E_2-E_3$ between the period-2 and -3 states at the Brillouin zone edge of the period-3 states, $P=P_{\rm edge}/3$. For the parameter of figure~\ref{fig_e}, $E_F/E_R=0.25$, we see that $E_2-E_3<0$ and thus the period-2 states are energetically lower than the period-3 states in the whole Brillouin zone of the latter (see figure~\ref{fig_e}). This figure shows that the period-2 states are always energetically lower than the period-3 states in the region of $0.245 \lesssim E_F/E_R \lesssim 0.4$ for $s=1$ and $1/k_Fa_s=-1$.

\bigskip
\noindent{\bf iii) Comments about rational and irrational number periods}

In principle, multiple period states with rational number periods are covered by our calculations with a supercell. Specifically, using a supercell with period $\nu$, we can describe period-$(\nu/\eta)$ states, where $\nu$ and $\eta$ are natural numbers. On the other hand, states with irrational number periods are excluded, which are beyond the scope of the present study. In our numerical calculations, which cover multiple period states with rational number periods, neither states with $\nu<\eta$ nor states whose period is incommensurate with the lattice constant appear as the lowest energy state. Therefore, it is probable that the multiple period states with irrational number periods could not be the energetically minimum states.

\bigskip
\noindent{\bf iv) Detailed information of the numerical calculations}

We set the parameters for the numerical calculations such as the number of grid points depending on the system parameter values to ensure the convergence. Here we provide detailed parameter values for the numerical calculations for $s=1$ and $E_F/E_R=0.25$ as a typical example.

In the transverse directions, we impose periodic boundary conditions with a large box size with $L_\perp/d=24$. Regarding the calculations of the stationary BdG equations, the cutoff energy $E_c$ we use is, for example, $E_c/E_F=40$ for $1/k_Fa_s=-1$, $40$--$200$ (mainly $100$) for $1/k_Fa_s=0$, and $100$--$200$ for $1/k_Fa_s=0.5$. The number of the grid points for $k_z$ within the first Brillouin zone $-q_B/\nu\le k_z \le q_B/\nu$ for period $\nu$ is $100$, $100$, $75$, $50$, and $40$ for $\nu=1$, $2$, $3$, $4$, and $5$, respectively. The number of the grid points in a supercell in the $z$ direction, $-\nu/2 \le z/d \le \nu/2$, is $200$ for $\nu=1$ and $100\nu$ for the other values of $\nu$. Regarding the TDBdG simulations shown in figure~\ref{fig_evolution}, the time discretization $\Delta t$ is $0.00008 \hbar/E_F = 0.00032 \hbar/E_R$ for $1/k_Fa_s=-1$ (left panel), $0.00005 \hbar/E_F = 0.0002 \hbar/E_R$ for $1/k_Fa_s=0$ (middle panel), and $0.00004 \hbar/E_F = 0.00016 \hbar/E_R$ for $1/k_Fa_s=0.5$ (right panel). Discretization of $k_z$ and $z$ are $\Delta k_z/q_B= 0.025$ and $\Delta z/d = 0.025$, respectively. Throughout the time evolution of figure~\ref{fig_evolution}, the total number of particles is conserved perfectly within the significant digits and the energy is conserved within $0.045$\% for $1/k_Fa_s=-1$ (left panel), $0.74$\% for $1/k_Fa_s=0$ (middle panel), and $2.2$\% for $1/k_Fa_s=0.5$ (right panel).


\section*{References}
\begin{thebibliography}{99}

\bibitem{soliton} Dauxois T and Peyrard M 2006 {\it Physics of Solitons} (New York; Cambridge University Press)

\bibitem{gruner} Gr\"uner G 1994 {\it Density Waves in Solids} (Massachusetts; Addison Wesley)

\bibitem{fflo} Fulde P and Ferrell R A 1964 {\it Phys. Rev.} {\bf 135} A550\\ \hspace{-1em}Larkin A I and Ovchinnikov Yu N 1965 {\it Sov. Phys. JETP} {\bf 20} 762

\bibitem{pasta} Watanabe G, Sonoda H, Maruyama T, Sato K, Yasuoka K and Ebisuzaki T 2009 {\it Phys. Rev. Lett.} {\bf 103} 121101\\ \hspace{-1em}Watanabe G and Maruyama T 2012, in {\it Neutron Star Crust} Chap. 2, pp. 23-44, eds. C. A. Bertulani and J. Piekarewicz (New York; Nova Science Publishers) (arXiv:1109.3511)

\bibitem{halo} Tanihata I, Hamagaki H, Hashimoto O. Shida Y, Yoshikawa N, Sugimoto K, Yamakawa O, Kobayashi T and Takahashi N 1985 {\it Phys. Rev. Lett.} {\bf 55} 2676

\bibitem{pitaevskii} Pitaevskii L P 1984 {\it JETP Lett} {\bf 39} 511

\bibitem{wu_st} Wu B, Diener R B and Niu Q 2002 {\it Phys. Rev. A} {\bf 65} 025601

\bibitem{diakonov} Diakonov D, Jensen L M, Pethick C J and Smith H 2002 {\it Phys. Rev. A} {\bf 66} 013604

\bibitem{mueller} Mueller E J 2002 {\it Phys. Rev. A} {\bf 66} 063603

\bibitem{machholm} Machholm M, Pethick C J and Smith H 2003 {\it Phys. Rev. A} {\bf 67} 053613

\bibitem{BEC_Kronig} Seaman B T, Carr L D and Holland M J 2005 {\it Phys. Rev. A} {\bf 71} 033622\\ \hspace{-1em}Seaman B T, Carr L D and Holland M J 2005 {\it Phys. Rev. A} {\bf 72} 033602

\bibitem{danshita} Danshita I and Tsuchiya S 2007 {\it Phys. Rev. A} {\bf 75} 033612

\bibitem{swallowtail} Watanabe G, Yoon S and Dalfovo F 2011 {\it Phys. Rev. Lett.} {\bf 107} 270404\\ \hspace{-1em}Watanabe G and Yoon S 2013 {\it JKPS} {\bf 63} 839

\bibitem{eckel} Eckel S, Lee J G, Jendrzejewski F, Murray N, Clark C W, Lobb C J, Phillips W, Edwards M and Campbell G K 2014 {\it Nature} {\bf 506} 200

\bibitem{lattice} Morsch O and Oberthaler M 2006 {\it Rev. Mod. Phys.} {\bf 78} 179

\bibitem{pethick_smith} Pethick C J and Smith H 2008 {\it Bose-Einstein Condensation in Dilute Gases, 2nd ed.} (New York; Cambridge University Press)

\bibitem{machholm04} Machholm M, Nicolin A, Pethick C J and Smith H 2004 {\it Phys. Rev. A} {\bf 69} 043604 

\bibitem{maluckov} Maluckov A, Gligori\'c G, Had\v zievski L, Malomed B A and Pfau T 2012 {\it Phys. Rev. Lett.} {\bf 108} 140402

\bibitem{gezerlis} Gezerlis A and Carlson J 2008 {\it Phys. Rev. C} {\bf 77} 032801(R)\\ \hspace{-1em}Gezerlis A and Carlson J 2008 {\it Phys. Rev. C} {\bf 81} 025803

\bibitem{vc_crossover} Watanabe G, Dalfovo F, Pitaevskii L P and Stringari S 2011 {\it Phys. Rev. A} {\bf 83} 033621

\bibitem{bloch} Bloch I, Dalibard J and Zwerger W 2008 {\it Rev. Mod. Phys.} {\bf 80} 885

\bibitem{giorgini} Giorgini S, Pitaevskii L P and Stringari S 2008 {\it Rev. Mod. Phys.} {\bf 80} 1215

\bibitem{degennes} de Gennes P G 1966 {\it Superconductivity of Metals and Alloys} (New York; Benjamin), chapter 5, p.~137

\bibitem{randeria} Randeria M in {\it Bose Einstein Condensation} ed Griffin A, Snoke D and Stringari S (Cambridge: Cambridge University Press) chapter 15, p. 355

\bibitem{bruun} Bruun G, Castin Y, Dum R and Burnett K 1999 {\it Eur. Phys. J. D} {\bf 7} 433

\bibitem{bulgac02} Bulgac A and Yu Y 2002 {\it Phys. Rev. Lett.} {\bf 88} 042504

\bibitem{grasso} Grasso M and Urban M 2003 {\it Phys. Rev. A} {\bf 68} 033610

\bibitem{optlatunit} Watanabe G, Orso G, Dalfovo F, Pitaevskii L P and Stringari S 2008 {\it Phys. Rev. A} {\bf 78} 063619

\bibitem{vc} Watanabe G, Dalfovo F, Piazza F, Pitaevskii L P and Stringari S 2009 {\it Phys. Rev. A} {\bf 80} 053602

\bibitem{miller} Miller D E, Chin J K, Stan C A, Liu Y, Setiawan W, Sanner C and Ketterle W 2007 {\it Phys. Rev. Lett.} {\bf 99} 070402

\bibitem{solitonlat} Buzdin A I and Tugushev V V 1983 {\it Sov. Phys. JETP} {\bf 58} 428\\ \hspace{-1em}Machida K and Nakanishi H 1984 {\it Phys. Rev. B} {\bf 30} 122

\bibitem{fedichev04} Fedichev P O, Bijlsma M J and Zoller P 2004 {\it Phys. Rev. Lett.} {\bf 92} 080401

\bibitem{orso05} Orso G, Pitaevskii L P, Stringari S and Wouters M 2005 {\it Phys. Rev. Lett.} {\bf 95} 060402

\bibitem{code} Bulgac A and Yoon S 2009 {\it Phys. Rev. Lett.} {\bf 102} 085302

\bibitem{ring} Ring P and Schuck P 1980 {\it The Nuclear Many-Body Problem} (New York; Springer)

\bibitem{regal} Regal C A, Greiner M and Jin D S 2004 {\it Phys. Rev. Lett.} {\bf 92} 040403

\bibitem{horikoshi} Horikoshi M 2014 private communication.

\end{thebibliography}
\end{document}